\documentclass[intlimits,twoside,a4paper]{article}
\usepackage{amsmath,amsfonts,amssymb}
\usepackage[T2A]{fontenc}
\usepackage[cp1251]{inputenc}
\usepackage{wrapfig}

\usepackage{cmpj2}

\issue{2014}{17}{4}{44001}
\doinumber{10.5488/CMP.17.44001}

\articletype{Rapid Communication}

\title[Morphologies for decorated metaparticles]
{Novel morphologies for laterally decorated metaparticles: Molecular dynamics simulation}

\author[A.Y.~Slyusarchuk, J.M.~Ilnytskyi]{A.Y.~Slyusarchuk\refaddr{label1}, J.M.~Ilnytskyi\refaddr{label2}}
\addresses{
\addr{label1} National University ``Lviv Polytechnic'', 12 Bandera St., 79013 Lviv, Ukraine
\addr{label2} Institute for Condensed Matter Physics of the National Academy of Sciences of Ukraine, \\ 1 Svientsitskii St., 79011 Lviv, Ukraine
}

\date{Received October 10, 2014, in final form December 4, 2014}
\authorcopyright{A.Y.~Slyusarchuk, J.M.~Ilnytskyi, 2014}

\newcommand{\idx}[1]{_{\mathrm{#1}}}

\newcommand{\fs}{~\textrm{fs}}

\newcommand{\nm}{~\textrm{nm}}

\newcommand{\Nch}{N_{\mathrm{ch}}}

\begin{document}

\maketitle

\begin{abstract}

We consider a mesoscale model for nano-sized metaparticles (MPs) composed of a central sphere decorated
by polymer chains with laterally attached spherocylinder. The latter mimics the mesogenic
(e.g., cyanobiphenyl) group. Molecular dynamics simulations of $100$ MPs reveal the existence of
two novel morphologies: $\textrm{uCol}_\mathrm{h}$ (hexagonal columnar arrangement of MPs with strong uniaxial order of
mesogens collinear to the columns axis) and $\mathrm{wCol}_\mathrm{h}$ [the same arrangement of MPs but with weak or no liquid crystalline (LC) order]. Collinearity of the LC director and the columnar axis in $\textrm{uCol}_\mathrm{h}$ morphology
indicates its potentially different opto-mechanical response to an external perturbation as compared to the
columnar phase for the terminally attached mesogens. Preliminary analysis of the structures of both phases is
performed by studying the order parameters and by visualisation of the MPs arrangements. Different mechanisms
for the mesogens reorientation are pointed out for the cases of their terminal and lateral attachment.


\pacs 02.70.Ns, 61.30.Vx, 61.30.Cz, 61.30.Gd
\end{abstract}

\section{Motivation}\label{I}

Colloid particles, polymers and LC molecules represent main building blocks of soft matter physics \cite{KlemLavrBook}.
When two or more of such blocks are combined into a MP, the latter forms a meta-material that exhibits a range of new
morphologies and new effects that are not observed for any of its pure constituents. Examples to mention are:
LC elastomers, LC dendrimers, decorated nanoparticles and others \cite{LC_elast,LC_elast.2,LC_dendr,LC_dendr.2,LC_gold,LC_gold.2},
all of these having already found a number of applications in thermo- and photo-controlled elasticity, plasmonic resonance, photonics
and medicine. Most applications rely on the symmetry of the equilibrium morphology, in which each constituent is
``responsible'' for particular property of a meta-material. For instance, its elasticity is usually controlled by a
polymer subsystem, optical properties are governed by both the behaviour of mesogens and the arrangement of gold nanoparticles
(if any). Therefore, the type of mutual arrangement of the constituent parts is crucial regarding the  new potential applications
of each particular meta-material.

\begin{wrapfigure}{i}{0.6\textwidth}
\begin{center}
\vspace{-3mm}
\includegraphics[clip,width=0.58\textwidth]{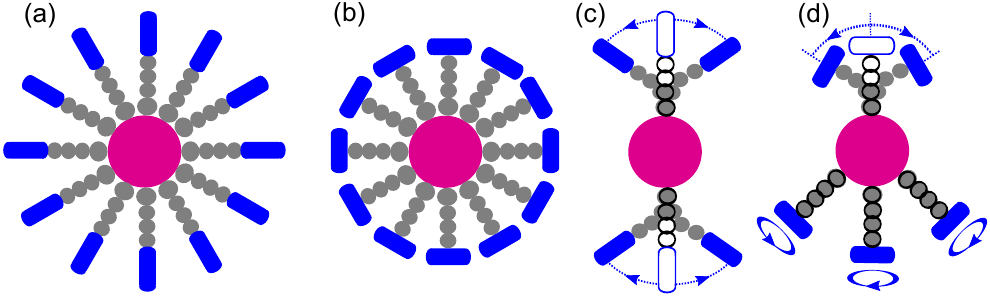}
\caption{\label{MP_model}(Color online) Schematic representation of a MP made of a spherical
core (shown in pink) decorated via spacers (shown in gray) each ending by a mesogen
(shown in blue). (a) MP with terminal attachment of mesogens, (b) the same with lateral attachment, (c) and (d):
reorientation options for attached mesogens for both models (see text for more details).}
\vspace{-4mm}
\end{center}
\end{wrapfigure}

Let us concentrate on the MPs built out of a spherical core decorated by polymer chains (spacers) each
ending by a mesogen. The core mimics either a solid nanoparticle (e.g., gold nanoparticle \cite{LC_gold,LC_gold.2}) or
averaged in time shape of a dendritic scaffold \cite{LC_dendr,LC_dendr.2}. It was found experimentally that the most important
aspects of decoration are as follows: the surface density of mesogens on the outer shell of MP,
the length of a spacer, and the exact way mesogens are attached chemically to the spacer \cite{LC_dendr,LC_dendr.2,LC_gold,LC_gold.2,LC_lateral,LC_lateral.2,LC_lateral.3,LC_lateral.4}.
In particular, both terminal and lateral attachment can be realised chemically \cite{LC_dendr,LC_dendr.2,LC_lateral,LC_lateral.2,LC_lateral.3,LC_lateral.4}
[depicted schematically in figure~\ref{MP_model}~(a) and (b)] and the difference between the morphologies observed in these two
cases are in the focus of this study.

\begin{wrapfigure}{i}{0.6\textwidth}
\begin{center}
\vspace{-3mm}
\includegraphics[clip,width=0.58\textwidth]{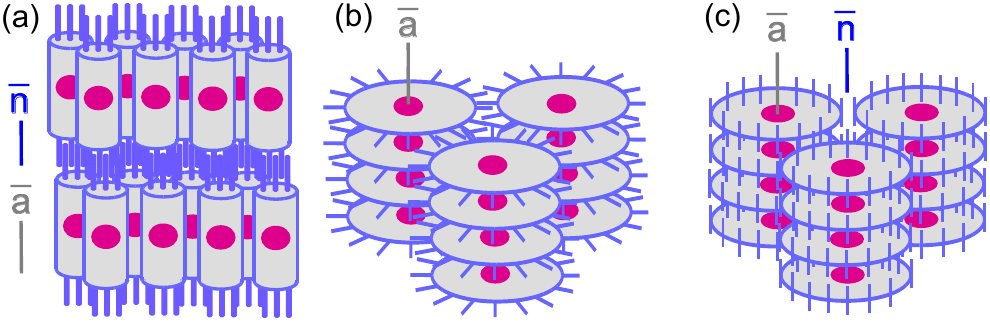}
\caption{\label{MP_phases_schema}(Color online) (a) and (b): schematic representation of lamellar smectic and
hexagonal columnar morphology, respectively, observed for the MP decorated by polymer chains with terminally
attached mesogens. (c): uniaxial hexagonal columnar morphology that is expected to exist for the case of
laterally attached mesogens.}
\vspace{-3.7mm}
\end{center}
\end{wrapfigure}

The case of a terminal attachment of mesogens [figure~\ref{MP_model}~(a)] has been studied more widely out of the
two \cite{LC_dendr,LC_dendr.2,LC_gold,LC_gold.2}. In particular, at low grafting density of chains such MPs adopt a rod-like
conformation
%
and self-assembly into lamellar smectic A morphology [see, figure~\ref{MP_phases_schema}~(a)].
With an increase of a grafting density, the disc-like conformation for MPs is favoured over the rod-like one, and the MPs
self-assemble into columns which arrange themselves hexagonally [see, figure~\ref{MP_phases_schema}~(b)].
Both structures possess uniaxial symmetry denoted via the symmetry axis $\bar{a}$. For the morphology (a) it is
defined as a normal vector to the layers (and it also coincides with the preferential direction for MPs rods),
whereas for its (b) counterpart $\bar{a}$ is defined along the columns of MPs. Despite this common feature, the
different type of MPs arrangement in these morphologies will affect their elastic properties and their behaviour
in plasmonic resonance applications (if the cores represent gold nanoparticles). Their behaviour will differ even
more in the applications related to the optical activity and orientational order of mesogens. Indeed, morphology
(a) is characterised by an uniaxial preferential orientation of mesogens defined via nematic director $\bar{n}$,
whereas for the case of morphology (b) the global director does not exist. The distribution of mesogens orientations
in this case is flat radial centered around the axes of each column [as indicated in figure~\ref{MP_phases_schema}~(b)].
One can clearly see some analogy with the case
of LC elastomers, where the role of a symmetry axis $\bar{a}$ can be attributed to the preferential orientation
of the polymer backbones \cite{LC_elast,LC_elast.2}. It is well known that their opto-mechanical applications depend heavily on the
mutual arrangement of $\bar{a}$ and $\bar{n}$ axes, the main- and side-chain architectures being an \linebreak example~\cite{LC_elast,LC_elast.2,LC_elast_photo,LC_our_simul,LC_our_simul.2,LC_our_simul.3,LC_our_simul.4}.

%
Therefore, a question arises whether some other morphologies can be obtained for the MPs with lateral attachment
of mesogens [figure~\ref{MP_model}~(b)] which are characterised by different mutual arrangement of the $\bar{a}$
and $\bar{n}$ axes, leading potentially to new optical and opto-mechanical behaviour. Based on this MP geometry,
one would expect the formation of hexagonal columnar morphology depicted in figure~\ref{MP_phases_schema}~(c). It should be
quite similar to its counterpart (b) (shown in the same figure) in terms of its mechanical properties and the arrangement
of cores, but, crucially, is characterised by uniaxial orientational order with the nematic director shown as
$\bar{n}$ therein. Moreover, are there any other morphologies possible? We try to provide the answer in the
following section by means of coarse-grained molecular dynamics simulations.

\section{Results and discussion}\label{II}

The model for a MP with laterally attached mesogens is based on its counterpart with terminal attachment
described in detail in references~\cite{ILW_dendr,ILW_dendr.2}. The central sphere represents a core
(be it a gold nanoparticle or a dendritic core) decorated by grafting $\Nch=32$ chains to it. We use the
annealing-like grafting, when the first bead of each chain slides freely on the surface of a central sphere.
The effective diameters of soft beads are based on the coarse-graining of the atomistic model for the generation
3 LC dendrimer performed in reference~\cite{HWS05}. These are: $2.14\nm$ for a central sphere, $0.62\nm$ for the
first bead and $0.46\nm$ for the remaining beads of the spacer which connects mesogens to the central sphere.

The spheres interact via soft repulsive
potentials of a quadratic form \cite{ILW_dendr,ILW_dendr.2}. The mesogens are modelled as soft spherocylinders of the
breadth $D=0.374\nm$ and of the length-to-breadth ratio $L/D=3$. These dimensions are also used for the
visualization purpose. The MP is kept together by means of harmonic spring potentials, with the same parameters
used for the model with terminal attachment \cite{ILW_dendr,ILW_dendr.2}. As far as each monomer is assumed to represent
three hydrocarbons \cite{HWS05} and their dimension is less than the Kuhn segment length, we introduce the
pseudo-valent angles which account for the effective spacer stiffness~\cite{ILW_dendr,ILW_dendr.2}.

Orthogonal lateral attachment of the mesogen to the spacer [figure~\ref{MP_model}~(b)] is maintained by employing
two potentials. First one
is the usual harmonic bond potential between the center or the last monomer of the spacer and the center of the
spherocylinder, the bond length is $0.3\nm$. The second potential is also of a harmonic type with respect to the
angle between the latter bond and the long axis of a mesogen \cite{Wils1997}. The equlibrium value for
this angle is set equal to $\theta_0=\pi/2$. The force constant for the latter potential is set equal to its
counterpart for the pseudo-valent angle potential mentioned above (the absolute value can be found in
references~\cite{ILW_dendr,ILW_dendr.2}).

In this study, we consider the MPs with $\Nch=32$ grafted chains only. The initial configuration is build out of
$100$ such MPs arranged randomly in a simulation box. The initial orientation of grafted chains is radial, with respect
to the central sphere. Typical box dimensions are of the order of $15\div20\nm$ and the periodic boundary
conditions are used in the simulations. First, a short run is performed of $10000$ molecular dynamics steps with a small time step of
$\Delta t=2\fs$ to eliminate initial interparticle overlaps. Following the scheme employed in references~\cite{ILW_dendr,ILW_dendr.2},
we performed field-assisted self-assembly simulations first. These are done at a fixed temperature $T=520~\mathrm{K}$ by
imposing an external orienting field. The latter is introduced via additional energy term
$U^{\mathrm{rot}}=-\sum^N_{i=1}{F\, \cos^2{\vartheta}_i}$,
which contains a contribution from each $i$th mesogen. Here, $\vartheta_i$ is the angle between the field direction and the orientation of $i$th
mesogen, $F$ is the magnitude of the field. Throughout this study we assume that the field is always directed along the $Z$ axis of the
Cartesian frame. Its magnitude $F$ sets a time-scale for the mesogens reorientation and we found empirically the values
in the range of $F=(2\div4)\cdot 10^{-20}~\textrm{J}$ to be optimal for our state points. The simulations are performed
in anisotropic isobaric ensemble, where the temperature and each of the three diagonal components of the stress
tensor are maintained constant by means of the thermostat and three separate barostats \cite{LC_our_simul,LC_our_simul.2,LC_our_simul.3,LC_our_simul.4}.
This makes possible the auto-adjustment of the simulation box shape which at the end becomes commensurate with the pitch of
the emerging ordered phase. One should remark that the potential interactions used in this model are predominantly soft
repulsive (save for the mesogen-mesogen potential introduced by Lintuvuori and Wilson \cite{LW08}). Therefore,
to maintain the density required for the ordered phases, the system needs to be stabilized by the pressures of
$P\geqslant 50~\textrm{atm}$ \cite{ILW_dendr,ILW_dendr.2}.

The field-assisted self-assembly simulations are performed for the range of pressures $P=50\div200~\textrm{atm}$, the
duration of each run is $10^{6}$ molecular dynamics steps with a time step of $\Delta t=20\fs$. For all values of the pressure within
this interval, the system indeed self-assembles into the uniaxial hexagonal columnar morphology depicted schematically in
figure~\ref{MP_phases_schema}~(c) and thereafter referred  to in this study as $\textrm{uCol}_\mathrm{h}$. Dependence of this morphology
on the pressure is reflected only in the density of stacking of MPs into columns, as well as in the total density of the system.
We will provide simulation snapshots for this morphology below. The next stage is to test the stability of this structure
and the changes it undergoes in a wide range of temperatures when the aiding field is removed ($U^{\mathrm{rot}}=0$). To provide a
quantitative analysis of these changes, let us introduce a set of the order parameters.

The level of orientational order of mesogens is described by the nematic order parameter defined as:
\begin{equation}\label{S_N}
  S_\textrm{N}=\left\langle P_2(\cos{\theta}_i)\right\rangle_{i,t}\,,
\end{equation}
where $\theta_i$ is the angle between the orientation of $i$th mesogen and nematic director axis $\bar{n}$, and
$P_2(x)=\frac{1}{2}(3x^2-1)$ is the second Legendre polynomial. The averaging is performed over all mesogens in
a system and over the time trajectory denoted as $\langle \ldots \rangle_{i,t}$. The initial director was set along the $Z$ axis
in the course of a field-assisted run. And, similarly to the case of MPs with terminally attached
mesogens \cite{ILW_dendr,ILW_dendr.2}, we found a negligible drift of the director away from this axis for all temperatures
where the orientation order exists whatsoever. Therefore, in this study we assume $\theta_i$ to be the angle between
the orientation of $i$th mesogen and the $Z$ axis.

\begin{wrapfigure}{i}{0.6\textwidth}
\begin{center}
\vspace{-8mm}
\includegraphics[clip,width=0.56\textwidth]{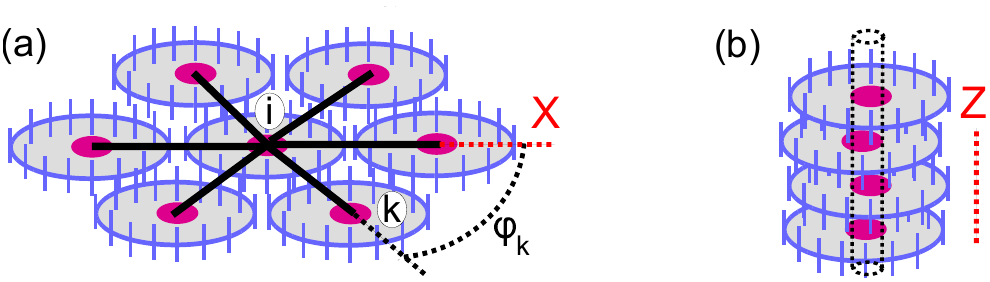}
\caption{\label{Hex_col_ordp}(Color online) (a) Flat hexagonal cluster of MPs and the definition of bond angles $\varphi_k$
 within it. (b) A column of MPs stacked along $Z$ axis.}
\vspace{-4mm}
\end{center}
\end{wrapfigure}
The amount of the ``hexagonality'' in the arrangement of MPs in the $XY$ plane can be characterized by the hexagonal order
parameter $S_\textrm{H}$, introduced as follows. Let us consider $i$th MP and identify its first coordination circle in the $XY$ plane.
To this end,
we evaluate the bond vectors $\mathbf{l}_{ik}=\mathbf{r}_{k}-\mathbf{r}_{i}=\{l^x_{ik},l^y_{ik},l^z_{ik}\}$
between its center to the centers of each of its neighbours indexed via $k$. The bond lengths between $i$th and $k$th MPs
projected onto the $XY$ plane are: $l^{xy}_{ik}=\big[{\big(l^x_{ik}\big)^2+\big(l^y_{ik}\big)^2}\big]^{1/2}$. The neighbour $k$ is assumed to belong
to the first coordination circle of $i$th MP if the following conditions fulfill: $R\idx{min}<l^{xy}_{ik}<R\idx{max}$ and
$|l^z_{ik}|<z\idx{max}$. For our model, we choose the following set of parameters: $R\idx{min}=4.7\nm$, $R\idx{max}=5.3\nm$
and $z\idx{max}=0.2\nm$ but the choice, obviously, depends on the geometry of the model MP. After the first coordination
circle that contains $N_k$ MPs are identified, the arbitrary axis in the $XY$ plane is chosen (e.g., the $X$ axis)
and the angles $\varphi_k$ are evaluated for each $k$th MPs belonging to this circle [see, figure~\ref{Hex_col_ordp}~(a)].
Then, both the local hexagonal order $S_{\textrm{H},i}$ for $i$th MP and its global counterpart $S_\textrm{H}$ can be defined via
equation~(\ref{S_H}) as:
\begin{equation}\label{S_H}
  S_{\textrm{H},i}=\left|\frac{1}{N_k}\sum_{k=1}^{N_k} \re^{6j\varphi_i}\right|\,, \qquad S_\textrm{H}=\langle S_{\textrm{H},i} \rangle_{i,t}\,, \qquad   S_{\textrm{C},i}=\frac{N_{\textrm{c},i}}{N\idx{max}}\,, \qquad S_\textrm{C}=\langle S_{\textrm{C},i} \rangle_{i,t}\,.
\end{equation}
Here, $j=\sqrt{-1}$.
The same equation also contains a definition for the other order parameter, $S_\textrm{C}$, which accounts for the ``columnarity''
of the $\textrm{uCol}_\mathrm{h}$ phase, i.e., the amount of stacking of the MPs discs along the $Z$ axis. It can be related to
the average columns height which is identified by considering $i$th MP and drawing through its center an imaginary cylinder
of radius $R_\textrm{c}$ that extends along the $Z$ axis [see, figure~\ref{Hex_col_ordp}~(b)].
Then, one takes into account the number
$N_{\textrm{c},i}$ of MPs so that their centers are found inside this cylinder (periodic boundary conditions are not taken into account).
Normalisation factor $N\idx{max}$ is introduced for convenience, to bring the range of $S_\textrm{C}$ values to the same interval as
the other two order parameters, $S_\textrm{N}$ and $S_\textrm{H}$. We used the following values: $R_\textrm{c}=1\nm$ and $N\idx{max}=15$.
These notations explain the expressions for the columnar order parameter $S_\textrm{C}$ in equation~(\ref{S_H}).

\begin{wrapfigure}{i}{0.6\textwidth}
\vspace{-4mm}
\begin{center}
\includegraphics[clip,width=.51\textwidth]{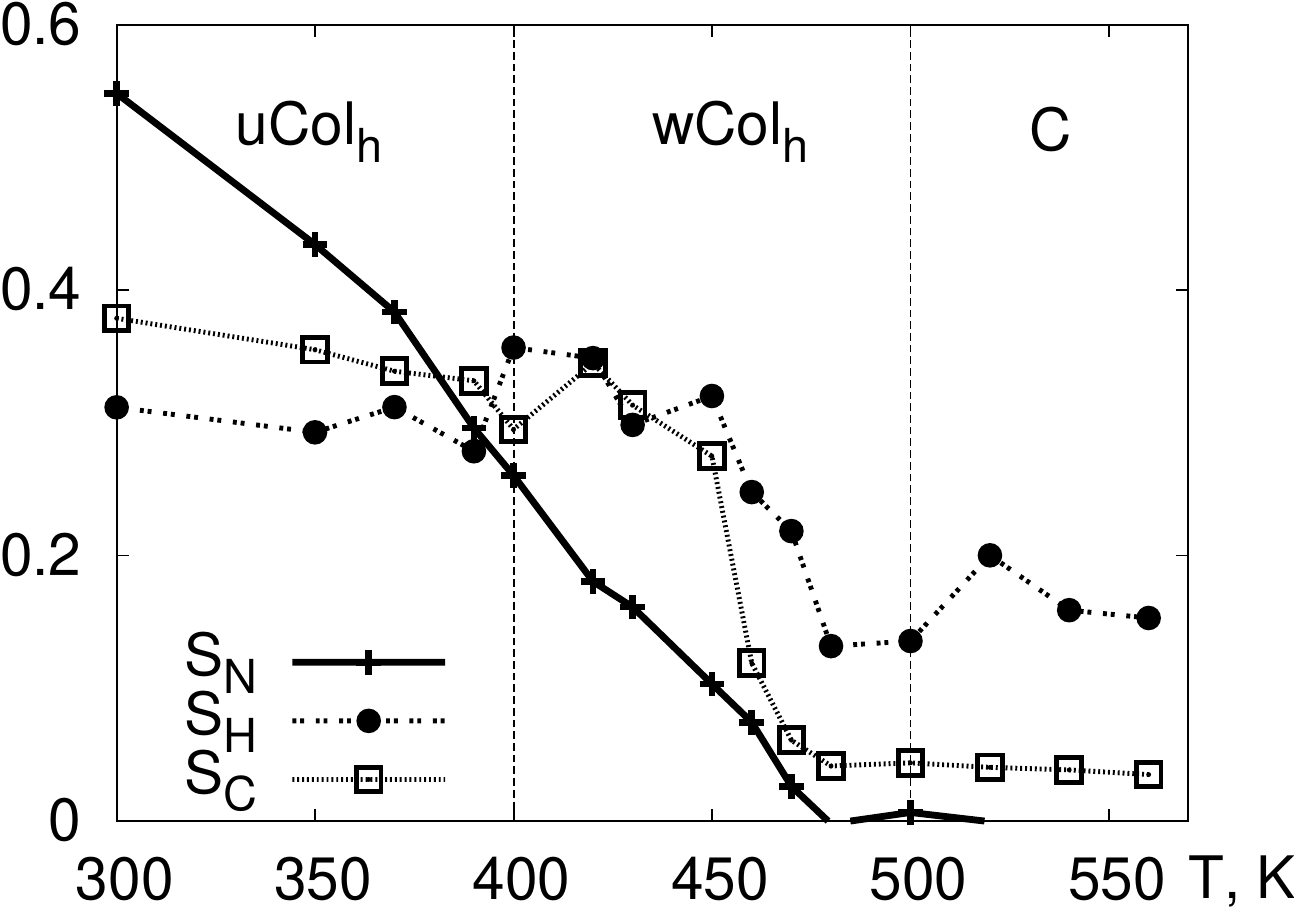}
\caption{\label{ord_params}Changes undergone by the order parameters $S_\textrm{N}$, $S_\textrm{H}$ and $S_\textrm{C}$ when $\textrm{uCol}_\mathrm{h}$
morphology, obtained by the field-aided self-assembly, is equilibrated at various temperatures from 300~K
to 560~K. The lines connect data points and serve as guides only.}
\end{center}
\vspace{-7mm}
\end{wrapfigure}
\begin{wrapfigure}{i}{0.6\textwidth}
\begin{center}
\vspace{-4mm}
\includegraphics[clip,width=0.58\textwidth]{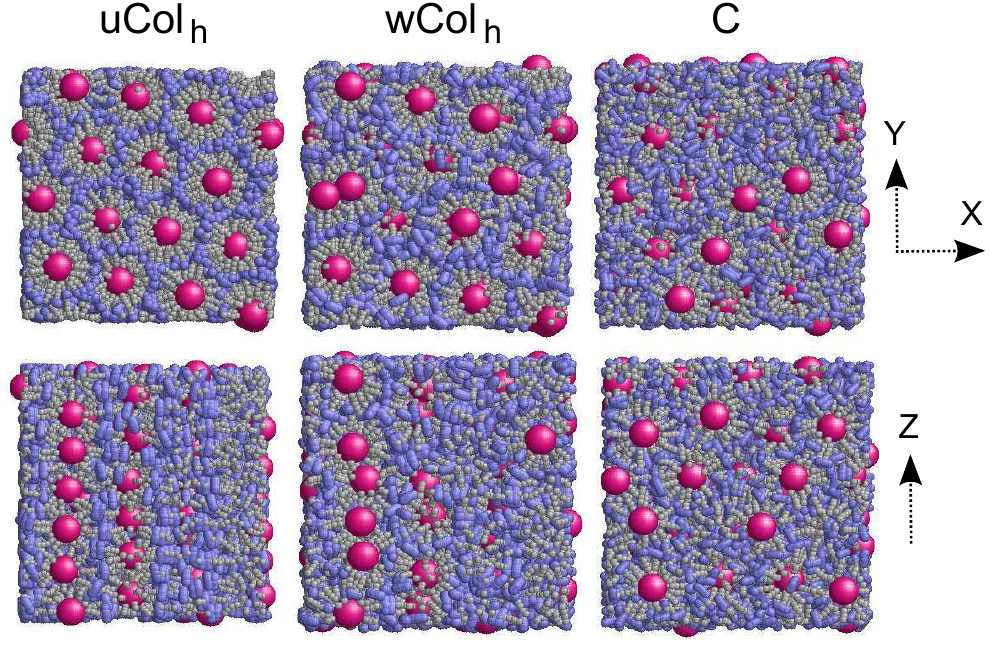}
\caption{\label{Snapshots_all}(Color online) Shapshots showing the structure of different morphologies.
Columns from left to right:  uniaxial hexagonal columnar at $T=300$~K ($\textrm{uCol}_\mathrm{h}$), weak hexagonal columnar
at $T=450$~K ($\mathrm{wCol}_\mathrm{h}$) and cubic at $T=520$~K (C) morphologies, all at the pressure of $100$~{atm}.
Rows from top to bottom: top view and side view for all respective cases (Cartesian axes are indicated in the figure).}
\vspace{-4mm}
\end{center}
\end{wrapfigure}

Let us concentrate on the behaviour of the order parameters $S_\textrm{N}$, $S_\textrm{H}$ and $S_\textrm{C}$ when the uniaxial
hexagonal columnar phase $\textrm{uCol}_\mathrm{h}$, obtained by the field-aided self-assembly, is equilibrated at various
temperatures. The plots demonstrating the changes are provided in figure~\ref{ord_params}. At relatively low
temperatures, $T<350$~K, the values of all three order parameters are essentially non-zero and one obtains the
uniaxial hexagonal columnar phase $\textrm{uCol}_\mathrm{h}$ first depicted schematically in figure~\ref{MP_phases_schema}~(c).
It is characterized by both uniaxial nematic order along
the $Z$ axis and regular columns of disc-shaped MPs aligned along the same axis and arranged hexagonally in the
$XY$ plane (see snapshots in the left-hand column of figure~\ref{Snapshots_all}). With an increase of the temperature,
all order parameters gradually decay to their minimum values. First of all, let us note that $S_\textrm{N}$ decays very
fast (practically linearly), in contrast to the case of terminally  attached mesogens \cite{ILW_dendr,ILW_dendr.2},
where the shape of the curve for $S_\textrm{N}$ had a semi-dome-like shape indicative of a power law form.
More importantly, in the case considered there, the decrease of $S_\textrm{N}$ and of the asphericity of the molecules
was synchronous: both turned to zero at about $T=500$~K. For the laterally attached mesogens considered here, we
observe an essential delay in the decrease of two other order parameters, $S_\textrm{H}$ and $S_\textrm{C}$, as compared to $S_\textrm{N}$. As one can
see in figure~\ref{ord_params}, there is a narrow temperature range around $T=450$~K, where the value of $S_\textrm{N}$ dropped
to about $0.1$ (typical of  the isotropic phase), but $S_\textrm{H}$ and $S_\textrm{C}$ are still almost the same as in the
$\textrm{uCol}_\mathrm{h}$ morphology at $T=300\div400$~K. One may refer to this morphology as a weak hexagonal columnar
($\mathrm{wCol}_\mathrm{h}$) and the important feature of it is that while it lost the orientation order of mesogens, its
hexagonal columnar structure is essentially preserved (see snapshots in the middle column of figure~\ref{Snapshots_all}).
One should note that the transition from $\textrm{uCol}_\mathrm{h}$ to $\mathrm{wCol}_\mathrm{h}$ is gradual and the boundary between
both shown in figure~\ref{ord_params} is rather for illustrative purpose. With further heating of the system, the
values for the order parameters $S_\textrm{H}$ and $S_\textrm{C}$ drop to their minima at approximately $T\geqslant 480$~K. The phase observed
at the temperatures higher than that is a cubic phase (C), where the MPs adopt the form close to spherical due to
an increased conformational entropy. Its structure is shown in the right-hand column of figure~\ref{Snapshots_all} and is the same as
for the case of terminally attached mesogens \cite{ILW_dendr,ILW_dendr.2}.

The faster decay of $S_\textrm{N}$ with an increase of the temperature and, as a result, the evidence of a new $\mathrm{wCol}_\mathrm{h}$
morphology can be explained by the following considerations. Let us go back to figure~\ref{MP_model}~(a) and (b) and consider
the difference in the allowed mechanisms for the mesogens reorientation in both cases. These are shown schematically in the
same figure, frames (c) and (d) for the cases of terminal and lateral attachment, respectively. For the case of a terminal
attachment, if we assume that the grafting point does not drift much, the only way for a mesogen to change its orientation
is to bend the spacer [see, figure~\ref{MP_model}~(c)]. There will be the energy penalty associated with this change due to
the pseudo-angle potential term which is introduced to model the stiffness of the alkyl chain at a meso-scale level. This
mechanism of bending the chain also exists for the case of laterally attached mesogens [figure~\ref{MP_model}~(d), top chain
with respect to the central sphere]. However, there also exists another mechanism of rotating a mesogen around the bond which
connects it to the end monomer of a spacer [figure~\ref{MP_model}~(d), bottom chains with respect to the central sphere].
This rotation costs no energy, as far as both bond length between a mesogen and the last monomer and the angle between
the orientation of a mesogen and the same bond stay unchanged. As a result, the rotational motion of mesogens in the mesogen-rich regions is more decoupled from the elastic energy of the spacers in the case of laterally attached mesogens
compared to the case of their terminal attachment. This is, most likely, the main reason for a faster decrease of the
nematic order with the raise of the temperature in the former case. We should note that the conformational changes in real
systems involve more factors to be considered, as compared to the current coarse-grained modelling. In particular, both
reorientation mechanisms depicted in figure~\ref{MP_model}~(d) are likely to be realized via soft interactions such as
torsional angles changes. Therefore, a more detailed analysis is needed to compare the energy penalty for both mechanisms
of the mesogens reorientation discussed here.

To conclude, we have shown the possibility of the formation of two novel morphologies, $\textrm{uCol}_\mathrm{h}$ and $\mathrm{wCol}_\mathrm{h}$
for the MPs with lateral attachment of mesogens. The former is uniaxial in both the symmetry of MPs packing and in
the arrangement of its mesogens. The latter is characterized by the same symmetry of MPs packing, while its LC subsystem is in the
isotropic phase. Both morphologies differ in their LC properties from their counterpart for the MPs with terminal attachment
of mesogens and are candidates for specific opto-mechanical applications.

\vspace{-6mm}


\ukrainianpart

\title{Нові морфології метачастинок декорованих шляхом бічного приєднання рідкокристалічних груп: компютерна симуляція}

\author{А.Ю.~Слюсарчук\refaddr{label1}, Я.М. Ільницький\refaddr{label2}}
\addresses{
\addr{label1} Національний університет ``Львівська політехніка'', вул. С.~Бандери, 12, 79013 Львів, Україна
\addr{label2} Інститут фізики конденсованих систем НАН України, вул. І.~Свєнціцького, 1, 79011 Львів, Україна
}

\makeukrtitle

\begin{abstract}
Розглянуто мезоскопічну модель нанорозмірних метачастинок (МЧ), які складається із центральної сфери, декорованої полімерними ланцюжками із бічним приєднанням сфероциліндрів. Останні описують мезогенні
(напр. цианобіфенільні) групи. Виконані симуляції $100$~МЧ за допомогою молекулярної динаміки вказують на існування двох нових морфологій: $\textrm{uCol}_\mathrm{h}$
(із гексагональним стовпцевим впакуванням МЧ та сильним одновісним впорядкуванням рідкокристалічних груп
колінеарно до осі стовпців) та $\mathrm{wCol}_\mathrm{h}$ (із тим же впорядкування МЧ але слабким або відсутнім
рідкокристалічним впорядкуванням). Колінеарність директора та осі симетрії стовпцевої фази у морфології
$\textrm{uCol}_\mathrm{h}$ вказує на її потенційно іншу оптико-механічну реакцію на зовнішнє збурення порівняно із
стовпцевою фазою для моделі із кінцевим приєднанням рідкокристалічних груп. Попередній аналіз структур обох
фаз виконано шляхом дослідження параметрів порядку та візуалізацією розташування МЧ. Вказано на роль різних
механізмів реорієнтації мезогенів залежно від способу їх приєднання.

\keywords наночастинки, рідкі кристали, самовпорядкування, молекулярна динаміка
\end{abstract}

\end{document}